\documentclass[12pt]{article}

\newtheorem{Theorem}{Theorem}[section]
\newtheorem{Definition}[Theorem]{Definition}
\newtheorem{Proposition}[Theorem]{Proposition}

\makeatletter
\usepackage{latexsym}
\usepackage[T1]{fontenc}
\usepackage{amsmath}
\usepackage{amssymb}

\newtheorem{Corollary}[Theorem]{Corollary}

\def \AO {{\cal A}({\cal O})}
\def \AO' {{\cal A}({\cal O}')}

\def \Pf {{\bf Proof.\,\,}}

\def \be {\begin{equation}}
\def \ee {\end{equation}}

\def \ume {{\scriptstyle{\frac{1}{2}}}}
\def \ra {\rightarrow}

\def \eqq {\equiv}

\def \a {{\alpha}}

\def \eps {{\varepsilon}}

\def \l {{\lambda}}
\def \La {{\Lambda}}
\def \s {{\sigma}}

\def \ph {{\varphi}}

\def \A {{\cal A}}
\def \B {{\cal B}}
\def \C {{\cal C}}

\def \G {{\cal G}}
\def \H {\mbox{${\cal H}$}}

\def \L {{\cal L}}

\def \M {{\cal M}}

\def \O {{\cal O}}

\def \S {{\cal S}}

\def \Z {{\cal Z}}

\def \id {{\bf 1 }}

\def \Rbf {{\bf R}}
\def \Cbf {{\bf C}}

\newfam\Ssfam
\font\eleSs=cmss10 at12pt \font\sevenSs= cmss10 at 8pt \font\sixSs= cmss10 at 6pt

\textfont\Ssfam=\eleSs\scriptfont\Ssfam=\sevenSs%
\scriptscriptfont\Ssfam=\sixSs \def\Ss{\fam\Ssfam\eleSs}

\def\doppio#1{{\rm I}\kern-.1667em{\rm #1}}

\def\Q{\text{Q}\kern-.52em
    \text{\vrule height1.5ex width.5pt depth0pt}\kern.45em}

\def\Z{{\mathchoice {\hbox{$\Ss\textstyle Z\kern-0.4em Z$}}
{\hbox{$\Ss\textstyle Z\kern-0.4em Z$}} {\hbox{$\Ss\scriptstyle Z\kern-0.25em
Z$}} {\hbox{$\Ss\scriptscriptstyle Z\kern-0.2em Z$}}}}

\def\C{{\mathchoice{\hbox{$\rm\textstyle\text{\kern.35em\vrule
   height1.5ex width.5pt depth0pt\kern-.35em C}$}}
{\hbox{$\rm\textstyle\text{\kern.35em\vrule
   height1.5ex width.5pt depth0pt\kern-.35em C}$}}
{\hbox{$\rm\scriptstyle\text{\kern.35em\vrule
   height1.5ex width.3pt depth0pt\kern-.35em C}$}}
{\hbox{$\rm\scriptscriptstyle\text{\kern.35em\vrule
   height1.5ex width.2pt depth0pt\kern-.35em C}$}}}}

\def \be{\begin{equation} \displaystyle}
\def \ee{\end{equation}}

\def \bea{\begin{eqnarray}}
\def \eea{\end{eqnarray}}

\def \Pf{{\em Proof.\,\,}}

\@addtoreset{equation}{section}

\def \be {\begin{equation} \displaystyle}

\def \ee {\end{equation}}

\def \cifm {C^\infty(\M)}
\def \difm  { \mbox{Diff}(\M)}
\def \gm {\G(\M)}
\def \gtm {\tilde{\G}(\M)}
\def \cmo {{C_0^\infty(\M)}}
\def \cif {{C^\infty}}
\def \co   {{C^\infty_0(\O)}}

\def \cst   {{C^*}}
\def \lm {{\cal L}(\M)}
\def \tv {T_v}
\def \lo {\L(\O)}
\def \ulv {U(\l v)}
\def \rv {R_v}
\def \rlv {R_{\l v}}
\def \rav {{R_{\a v}}}
\def \am {\A(\M)}

\def \ois {{\O \in \S}}
\def \glv  {{g(\l v)}}
\def \GM  {{\G(\M)}}
\def \Gtm  {{\tilde{\G}(\M)}}
\def \tcmo {{\tilde{C}^\infty_0(\M)}}
\def \CM  {\Pi(\M)}
\def \tilm {\tilde{\cal L}(\M)}
\def \ao {{\A(\O)}}

\begin{document}
\begin{titlepage}
\title{The Noncommutative Poisson Algebra of Classical and Quantum Mechanics}

\sloppy

\author{G. Morchio \\Dipartimento di Fisica, Universit\`a di Pisa,
\\and  INFN, Sezione di Pisa, Pisa, Italy \and F. Strocchi \\
Scuola Normale Superiore, Pisa, Italy \\ and INFN, Sezione di Pisa}

\fussy

\date{}

\maketitle

\begin{abstract}The Lie-Rinehart (LR) algebra of a manifold $\M$, defined by the Lie
structure of the vector fields, their action and their module structure on
$C^\infty(\M)$, is a common, diffeomorphism invariant, algebra for both
classical and quantum mechanics. Its (noncommutative) Poisson universal
enveloping algebra $\La_R(\M)$ contains a central variable $Z$ which relates
the commutators to the Lie products; classical and quantum mechanics are its
only factorial realizations, corresponding to $Z = i z$, $z = 0$ and $z =
\hbar$, respectively. In this form, canonical quantization appears therefore as
a consequence of such a general geometrical structure. The regular factorial
Hilbert space representations of $\Lambda_R(\M)$ are, for $z \neq 0$, unitarily
equivalent, apart from  multiplicity, to one of the irreducible quantum
representations, with $z =\hbar$, which are locally Schroedinger and in one to
one correspondence with the unitary irreducible representations of the
fundamental group of $\M$. For $z = 0$, if Diff($\M$) is unitarily implemented,
they are unitarily equivalent, up to  multiplicity, to the representation
defined by classical mechanics on $\M$.

\end{abstract}

\vspace{15mm} \noindent Math. Sub. Class.: 81S10, 81Q70, 81R15, 53D17, 81R60,
81R10

\vspace{3mm}  \noindent Key words: Quantum mechanics on  manifolds,
Lie-Rinehart algebras, Noncommutative Poisson algebras, Canonical quantization,
Classical mechanics

\end{titlepage}

\newpage
\section{Introduction}
In a previous work ~\cite{MS1} we discussed the formulation of Quantum
Mechanics (QM) on a manifold $\M$ emphasizing the role of diffeomorphism
invariance and of the associated Lie-Rinehart (LR) algebraic structure  of the
generators of $\difm$ as a module on $\cifm$. Such a construction relies on
structures which are associated  to the geometry of the (configuration)
manifold $\M$ and are also at the basis of classical mechanics. It is therefore
relevant to ask where and to which extent the classical and quantum approaches
differ, starting from such  common geometric structures.

In fact, we shall show that a (slight) generalization of the algebraic
framework of Ref.~\cite{MS1} gives rise to an algebraic structure which is
common to Classical (CM) and quantum mechanics and actually has no other
realizations. CM and QM will arise from the value of a central variable
intrinsically provided by the common algebraic structure. It is worthwhile to
stress that the symplectic geometry of the classical phase space is not an {\em
a priori} common  ingredient, but it  rather characterizes
 one of the values of
the central variable. Thus, the emerging structural relation between CM and QM,
far from treating the second as a deformation of the classical symplectic
structure, identifies both as realizations of the same algebra.

\def \lrm {{\cal L}_R(\M)}
\def \lr {{\cal LR}}
\def \LM {{\La_R(\M)}}
\def \lom {{C^\infty(\M)}}
\def \lam {{\La(\M)}}
The analysis relies on the intrinsic {\em LR algebraic structure} associated to
the {\em configuration manifold} $\M$ and to $\difm$ in terms of  i) the
algebra  $\cifm$ generated by the $C^\infty$ real functions on $\M$ with
compact support and by the constant functions, ii) the Lie algebra $\lm$ of
$\cif$ vector fields $v$ of compact support, with Lie product $\{ . \,, \,.\}$,
iii) the action of $\L(\M)$ on $\cifm$, denoted by $\{ v, \,f\}$ and iv) the
module structure of $\L(\M)$ over $\cifm$, given by the {\em LR product}: $
\forall f \in \cifm$, $ \forall v \in \lm$, $(f, v) \ra f \circ v \in \L(\M)$,
which is linear in both factors, associative in the first and satisfies $\{ f
\circ v, \,g\} = f \{v, \,g\}$.

The LR  algebra $ \lrm \eqq (\cifm, \lm)$  uniquely determines its non
commutative Poisson universal enveloping algebra $\La_R(\M)$ (see Section 2).
$\LM$ has a Lie product $\{.\,, \,.\}$, which extends that of $\lrm$ and an
{\em associative product}, denoted by $\cdot$\,, whose symmetric (Jordan) part
extends the partial product in $\lrm$.  The Lie product satisfies the Leibniz
rule with respect to the associative product. No relation is assumed between
the commutator $[\,A, \,B\,] \eqq A\cdot B - B \cdot A$ and the Lie product.

The {\em noncommutative Poisson algebra} $\LM$ uniquely defines a notion of
($\difm$ invariant) noncommutative geometry of ($\M$, $Vect(\M)$), only
embodying the LR relations. Such a noncommutative geometry underlies both CM
and QM; in fact $\LM$ is represented in both, with $[\,A,\,B\,] = 0$ in the
classical case, yielding the classical commutative Poisson algebra and the
symplectic structure of the phase space; in the quantum case  $[\,A, \,B\,] = i
\,\hbar \{ A, \,B \}$,  yielding a realization of $\LM $ through the Heisenberg
algebra, with the LR product  realized  as the symmetric (Jordan) product, as
recognized in Ref. ~\cite{MS1} (see Section 4).

The main point of our analysis is that $\LM$ contains a {\em central} variable
$Z$ (a central sequence if the manifold is not compact), which relates the
commutator to the Lie product: \be{[\,A, \,B\,] = Z \cdot \{ A,\, B\},
\,\,\,\,\,\,\,\,\,\,[\,A, \,Z\,] = 0 = \{ A, \,Z \}, \,\,\,\,\,\forall A, B \in
\LM.}\ee Under the standard reality assignment: $f^* = f$, $v^* = v$, $ \forall
f \in \cifm$, $ v \in \L(\M)$, $\LM$ becomes a Poisson *-algebra and relations
(1.1) imply that classical and quantum mechanics are its only ``factorial
realizations'', i.e. such that the central variable $Z$ takes a definite value
$\l \in \Cbf$ (Sections 3, 4).

This shows that the Dirac ansatz of canonical quantization, i.e. the
proportionality of the commutators of {\em positions and momenta} to their
classical Poisson brackets,  has no alternative, in the sense that the only
possibility for the commutator between variables in $\lrm$ is given by their
Lie product times a central variable. Hence, "quantization" is determined by
the LR structure of $\lrm$, the classical Poisson algebra and the Heisenberg
algebra resulting as the  unique alternatives.

Section 2 is devoted to the definition of  the LR structure of $\lrm$, to the
analysis of its localization properties and to the construction of $\LM$ as its
unique non commutative Poisson universal enveloping algebra.

In Section 3, we exploit  the localization properties of $\LM$ induced by the
LR local structure of $\lrm$. This allows for the construction  of variables
which are locally central with respect to both the Lie and the commutator
product. For a compact manifold they sum up to a (global) central variable $Z$,
which relates the commutator to the Lie product, eq.\,(1.1). For non compact
manifold, they give rise to a central sequence and therefore to an extension of
$\LM$ with a central variable $Z$ yielding eq.\,(1.1).

In Section 4, we consider  the regular  Hilbert space representations $\pi$ of
the Poisson algebra $\La_R(\M)$, characterized by exponentiability of the
representatives of the vector fields and of the central variable $Z$ to one
parameter  groups $U(\l v)$, $U(\l Z)$, and by strong continuity of $\difm$ as
automorphisms of the $C^*$-algebra  generated by $\pi(C^\infty(\M)) $ and by
the $U(\l v)$, $U(\l Z)$. Such representations are called {\em factorial} if
the $\difm$ invariant  elements of the center of the corresponding Von Neumann
algebra are multiple of the identity. We show that i) for $z \neq 0$, the
regular factorial Hilbert space representations of $\Lambda_R(\M)$ are
unitarily equivalent, apart from a multiplicity, to one of the irreducible
quantum representations, with $z =\hbar$, classified in ~\cite{MS1}, which are
locally Schroedinger and in one to one correspondence to the unitary
irreducible representation of the fundamental group of $\M$; ii) for $z = 0$,
if Diff($\M$) is unitarily implemented, they are unitarily equivalent, up to a
multiplicity, to the representation defined by classical mechanics on $\M$.

The above analysis shows the basic role of the LR geometry associated to the
configuration manifold $\M$, which in the quantum case   is somewhat hidden in
the $C^*$-algebraic structure (since the Lie product is given by the commutator
apart from the Planck constant), and in the classical case has nothing to do
with  the  (abelian) algebraic structure. In fact, the LR geometry of $\difm$
provides a general  notion of mechanical system, in terms of the Poisson
universal enveloping algebra $\LM$,  classical and quantum mechanics being
uniquely determined as its only factorial realizations. Such a strong
implication is simply displayed in the case of ${\bf R}^n$, with the
translation group playing the role of the diffeomorphisms.

This suggests a quite different approach to the relation between Classical and
Quantum mechanics, with respect to phase space ``quantization''. The latter
assumes as fundamental the classical canonical phase space with its symplectic
diffeomorphism group, and   associates $C^*$-algebras $\A_\hbar$ to it. Rather
than trying to obtain Quantum mechanics from the symplectic structure of the
classical phase space, our approach recognizes as fundamental the {\em
Lie-Rinehart noncommutative geometry of the configuration space $\M$ and of
$Vect(\M)$}, the classical symplectic space and the quantum mechanical state
space resulting from the central classification of the same noncommutative
Poisson algebra.

\vspace{1mm}
\def \mtre {\vspace{-3.5mm}}
We shall adopt the following {\bf Notations}: \vspace{1mm}\begin{description}
\mtre \item {$\M$} a connected $C^\infty$ manifold of dimension $d$, \mtre
\item{$\O$} any subset of $\M$ diffeomorphic to an open sphere, \mtre
\item{$\difm$} the connected component of the identity of the group of
diffeomorphisms of $\M$, \mtre
\item{$\lm, \, \lo$} the Lie algebras of $C^\infty$ vector fields $v$ of compact
support in $\M, \O$, with Lie product $\{v_1, \,v_2\}$, \mtre \item{$\glv$},
$\l \in \Rbf$, $v \in \lm$, the associated one parameter groups, which exist by
compactness of supp\,$v$, \mtre \item{$\gm$} the subgroup of $\difm$ generated
by the $\glv$, \mtre \item{$\gtm$} its universal covering group, which is
uniquely associated to $\lm$ (\cite{Mi}, Theorem 8.1) and  is generated by the
elements of a neighborhood of the identity in $\GM$ and therefore by the  one
parameters groups $\glv$, \mtre
\item{$C_0^\infty(\M) $, $C^\infty(\B)$} the algebras generated by the  $C^\infty$
real functions  with compact  support, respectively  in $\M$ and in the open
set  $\B$, \mtre
\item { $C^\infty(\M)$}  the  algebra generated
by  $C_0^\infty(\M)$ and by the constant functions.
\end{description}

%%%%%%%%%%%%%%%%%%%%%%%%%%%%%%%%%%%%%%%%%%%%%%%%%%%%%%%%%%%%%%
%%%%%%%%%%%%%%%%%%%%%%%%%%%%%%%%%%%%%%%%%%%%%%%%%%%%%%%%%%%%%%%%%%%%%

%%%%%%%%%%%%%%%%%%%%%%%%%%%%%%%%%%%%%%%%%%%%%%%%%%%%%%%%%%%%%%%%%%
\sloppy
\section{The Lie-Rinehart  noncommutative Poisson \\ algebra of $\M$} \fussy
We recall the generalization of the notion of Poisson algebra to the
noncommutative case~\cite{FGV}~\cite{Far}:
\begin{Definition} A {\bf Poisson algebra} $\Lambda$ is a  real associative
algebra, with product denoted by $A
\cdot B$, and a (real) Lie algebra, with product denoted by $\{A,\,B\}$,
satisfying  the Leibniz rule \be{\{A, \,B \cdot C\}= \{A, \,B\} \cdot C + B
\cdot \{A,\,C\},}\ee
\end{Definition}
Our analysis will be based on the following notion:
\begin{Definition}
  Given a (real) Lie algebra
$\L$, with Lie product $\{.\,,\,.\}$, a {\bf Poisson enveloping algebra}
$\La_\L$ of $\L$  is  a Poisson algebra with  a Lie algebra homomorphism $i: \L
\ra \La_\L$: $\forall l_1, \,l_2 \in  \L$, \be{ i( \{l_1, \,l_2\}) = \{ i(l_1),
\,i(l_2)\}.}\ee A Poisson enveloping algebra $\La_\L$ of a Lie algebra $\L$
will be called {\bf universal} if, for any Poisson enveloping algebra
$\La_\L'$, there is a unique homomorphism $\rho: \La_\L \ra \La_\L'$ of non
commutative Poisson algebras, which satisfies $i_{\La_\L'} = \rho \circ
i_{\La_\L}$.
\end{Definition}
As in general for enveloping algebras, the uniqueness of the  Poisson universal
enveloping algebra (UEA) of a Lie algebra $\L$ follows immediately from the
uniqueness of the homomorphism $\rho$. Its construction proceeds as in the
Poincaré-Birkoff-Witt theorem by considering the tensor product polynomial
algebra generated by $\L$ and by taking quotients with respect to the ideals
generated by  eqs.\,(2.1), (2.2). By construction $i$ is injective.

We  recall the notion of Lie-Rinehart algebra~\cite{R}: \begin{Definition} A
real {\bf Lie-Rinehart (LR) algebra} is a pair $(\L_0, \L)$, where $\L_0$ is a
real commutative algebra, $\L$ is a (real) Lie algebra, with Lie product
$\{.\,,\,.\}$, which acts as derivations on $\L_0$, $\forall f \in \L_0$, $v
\in \L$, $ f \ra v(f)$ (hereafter $v(f)$ is denoted also by $\{v, \,f\}$);
$(\L_0, \L)$ is equipped with a  (LR) product $\L_0 \circ \L \subset \L$,
satisfying  distributivity in both factors and, $\forall \,f, g \in \L_0$, $v,
w \in \L$ \be{ \{\,f\circ v,\,g\,\} = f\,\{v, \,g\},}\ee\be{f \circ(g \circ v)
= (f g) \circ v,}\ee\be{\{v, \,f \circ w\} = \{v,\,f\} \circ w + f \circ \{v,
\,w\}.}\ee A Lie-Rinehart algebra is said to have an identity, if the algebra
$\L_0$ has an identity $\id$, which automatically satisfies $ \{ \id, \, v \} =
0, \,\,\,\forall v \in \L$.
\end{Definition}
Clearly, $\L_0 + \L$ is a Lie algebra, with Lie product $\{.\,,\,.\}$ and we
define
\begin{Definition}
The {\bf Poisson universal enveloping algebra of a LR algebra $(\L_0, \L)$}
with identity is the Poisson universal enveloping algebra of the Lie algebra
$\L_0 + \L$, with the additional relations $\forall f, g \in \L_0$, $\forall v
\in \L$,
\newline i)  $i(\id) \cdot i(l) = i(l)$, $\forall\, l \in (\L_0 + \L)$
\newline ii)  $i(f g) = i(f)\, i(g)$,
\newline iii) $i(f \circ v) = \ume (i(f) \cdot i(v) + i(v) \cdot i(f))$.
\end{Definition}

A Poisson UEA  is not an  enveloping algebra  in the sense of envelops of Lie
or of LR algebras~\cite{R}, since no relation is assumed between the Lie
product and the commutator. Obviously, the Poisson UEA of a LR algebra is
isomorphic to the quotient of the Poisson UEA of $\L_0 + \L$ with respect to
the ideals  defined by  i)- iii). Existence and uniqueness of Poisson UEA
follow as before.

\def \alv {g_{\l v}}
\def \LM {\La_R(\M)}
$\lm$ acts as derivations on $C^\infty(\M)$ and $(\lom, \lm)$ is a LR algebra
with LR product defined by eq.\,(2.3). The same holds for
 $(C^\infty_0(\O), \L(\O))$. $C^\infty(\M)$ has an identity $ 1 = 1(x)$.
Clearly, $\difm$ defines a group of automorphisms of  $(\lom, \,\lm)$ and of
its Poisson UEA $\Lambda_R(\M)$. The relation between the action of $\G(\M)$ on
$\Lambda_R(\M)$ and the Lie product $\{.,\,.\}$ is \be{ (d/ d \l)\, g_{\l v}(A)
= \{\,v,\,\alv(A)\,\}, \,\,\,\,\,\forall A \in \LM,}\ee where $\alv$ denotes
the one parameter subgroup of $\difm$ generated by $v$ and the derivative is
taken in the $C^\infty$ topology.

\def \\L_0(\M) {C^\infty(\M)}
\begin{Definition} The
  {\bf LR noncommutative Poisson algebra} $\La_R(\M)$ of a manifold $\M$ is the Poisson  universal
enveloping algebra of the LR algebra $(C^\infty(\M), \L(\M))$.

We denote by $\La(\M)$ the Poisson UEA of $C^\infty(\M) + \L(\M)$ satisfying
only condition i) of Definition 2.4.
\end{Definition}
With respect to $\LM$, the Poisson UEA of $C^\infty(\M)+ \L(\M)$ merely
embodies the Lie relations between vector fields and functions on $\M$, so that
it applies in general to the analysis of $\difm$. In particular, such a Lie
algebra can be interpreted as the current algebra, which appears in all
$N$-particle systems on $\M$~\cite{G}. With respect to such a general
structure, as discussed in Ref.~\cite{MS1}, the mechanics of a particle on $\M$
requires the elimination of  redundant degrees of freedom, which is provided by
the LR relations between otherwise independent vector fields.

The algebra $\LM$ can be interpreted as describing a particle on a manifold
solely on the basis of general geometrical relations, which hold both in  the
quantum and in the classical case, since neither quantum commutation relations
(as
 in Ref.~\cite{MS1}), nor  the vanishing of the commutator are assumed .

An important property of $\LM$ is that there is a notion of localization in the
following sense: an element $A \in \LM$ has  the compact set $K$ as a {\em
localization region}, if it can be represented by  a polynomial of elements of
$\L_0(\B) + \L(\B)$, for some $\B \subset K$. In the case of non compact
manifold $\M$, $\La_R(\M)$ is generated by the identity and by localized
elements.
\begin{Proposition} If $A \in \LM$ is localized in the compact set  $K$ and $g \in C^\infty(\M)$
satisfies supp $g \cap K = \emptyset$, then \be{ i(g)\cdot A = A \cdot i(g) =
0.}\ee Therefore, by i), if  $g(x) = 1$\, $\forall x \in K$, then \be{
i(g)\cdot A = A \cdot i(g) = A.}\ee
\end{Proposition}
\Pf \,If $A \in C^\infty(\B)$, $\B \subset K$, the result follows from ii),
trivially.

\noindent If $A = v \in \L(\B)$,  for $ f \in C^\infty(\B)$, with $f(x) = 1 $,
$ \forall x \in$ supp $v$, one has $f \circ v = v$ and therefore, as a
consequence of ii), iii), $$i(g) \cdot i(v) = \ume i(g) \cdot [ (i(f) \cdot
i(v) + i(v) \cdot i(f)] = \ume \,i(g) \cdot i(v) \cdot i(f). $$ By iii), and
ii),  $ g \circ v = 0$ implies that the r.h.s. is equal to $ -\ume i(v) \cdot
i( g f ) =0$. By associativity of the product, eq.\,(2.7) extends to all $A \in
\LM$ localized in $K$. $\hfill \square$

\section{The relation between  commutators and  Lie
products }

As remarked before, in a Poisson algebra there is no {\em a priori} relation
between the commutator and the Lie product. As we shall prove in this section,
for the LR Poisson algebra $\LM$ the two products are not independent: for a
compact manifold $\M$, there is  a central variable which relates the
commutator to the Lie product by eq.\,(1.1). In the non compact case, there are
sequences which are central both in the Lie and in the commutator sense and
give the same relations. The existence of central relations between commutators
and Lie products has been recognized for  strictly non commutative Poisson
algebras which are prime, i.e. such that they do not have ideals which are
divisors of zero~\cite{Far}. This is not the case of $\LM$, since in particular
any pair of functions with disjoint supports generate (bilateral) ideals $I_1,
I_2$ with $I_1\cdot I_2 = 0$ (as a consequence of Theorem 3.1 below). For
brevity, in the following the injection $i$ will not be spelled out;  $[\,A,
\,B\,] \eqq A \cdot B - B \cdot A$ will denote the commutator.
\begin{Theorem} For a compact manifold $\M$, there exists a unique $Z \in \LM$,
such that, $\forall A, \,B \in \LM$, \be{ [\,A, \,B \,] = Z\,\cdot \{\,
A,\,B\,\}, \,\,\,\,\,\,\,\,\,\,\,\{\, Z,\, A\,\} = 0 = [\,Z,\,A\,].}\ee For a
non compact manifold $\M$, there exists a sequence $Z_n \in \LM$, such that $
\forall A, \,B \in \LM$, \be{ [\,A, \,B \,] = Z_n\,\cdot
\{\,A,\,B\,\},\,\,\,\{\, Z_n,\, A\,\} = 0 = [\,Z_n,\,A\,], \,\,\,\forall n
> \bar{n}(A,B).}\ee Then, one can define an element $Z$, such that the Poisson
algebra $\tilde{\La}_R(\M)$  generated by $\LM$ and $Z$ satisfies eq.\,(3.1).
\end{Theorem}
\Pf\, A crucial role is played by  the property of $\difm$, by which the linear
span of $\{ C^\infty(\M), \,\L(\M) \} $ contains $C^\infty(\B)$, $\B$ any open
set contained in a compact set $K$. In fact, for each open region $\O_i$,
diffeomorphic to a sphere, by choosing a sphere $\O_i'$ larger than $\O_i$, one
can find $q_i \in C^\infty(\O_i')$, $w_i \in  \L(\O_i') $ such that $\{ q_i,
\,w_i\}(x) = 1$, $\forall x \in O_i$.  Then, for any $g_i$ with support
contained in $\O_i$,  in the LR algebra $(C^\infty( M), \L(\M))$  one has\be{
g_i = g_i \{ q_i, \,w_i \} = \{\,q_i, \,g_i \circ w_i \} ,}\ee and, therefore,
any $g \in C_0^\infty(\M) $ has a decomposition  \be{g = \sum_i g_i = \sum_i\{
q_i, \,g_i \circ w_i \}, \,\,\,\, \,\,\,\,\,\mbox{supp}\,g_i \subset \O_i.}\ee

\noindent Then, by using the following identity, which holds in Poisson
algebras,~\cite{Far} $$ [\,A, \,B\,]\,\cdot\{\,C, \,D\, \} =
\{\,A,\,B\,\}\,\cdot [\,C,\,D\,]\,,$$  putting $ Z_g \eqq \sum_i [q_i, \,g_i
\circ w_i ]$, $\forall A,\,B \,\in \LM$, one obtains \be{ [\,A,\,B\,]\cdot g  =
\sum_i [ A,\,B ] \cdot \{q_i, \,g_i \circ w_i\} = \{A, \,B\} \cdot Z_g\,, }\ee
\be{ g \cdot [\,A,\,B\,] = Z_g \cdot \{\,A,\,B\,\}.}\ee The element $Z_g$ may
depend  on the choice of $g_i, q_i, w_i$ in the decomposition of $g$,
eq.\,(3.4); however, by the same identity,   for any $h$ of compact support one
has \be{ Z_g \cdot h = \sum_i \{\,q_i, \,g_i \circ w_i\} \cdot\,Z_h = g
\cdot\,Z_h = Z'_g \cdot h,}\ee where, for given $g$, $Z'_g$ corresponds to any
other choice in the above construction.\goodbreak

\noindent Furthermore, for any $A$ such that $ \{\,A, g\} =0$, in particular
for all $A$ such that $g(x) = 1$, $ \forall x \in $ supp\,$A$, putting for
brevity $p_i \eqq g_i \circ w_i$, using  Leibniz rule and eq.\,(3.5), one has
$$ \{[ q_i, \,p_i ], \,A\} \cdot\,g = ( [ \{q_i, \,A\}, p_i ]  + [ q_i,
\,\{\,p_i, \,A\} ]) \cdot g = $$ $$ = (\{\,\{q_i, \,A\}, p_i\,\} + \{\,q_i,
\,\{\,p_i, \,A\,\}\,\} \,)\cdot Z_g = \{\,\{q_i, \,p_i\}, \,A\}\,\cdot Z_g\,.
$$ Hence, \be{ \{ Z_g \cdot g, \,A\} = \sum_i \{[ q_i, \,p_i ], \,A\} \cdot\,g
= \{ g ,\,A\} \cdot Z_g = 0}\ee and \be{ [ Z_g \cdot g, \,A ]  = [\, Z_g,\, A ]
\cdot g = \{Z_g , \,A \} \cdot Z_g = 0.}\ee In the case of a compact manifold,
by taking $g = 1$, condition i) implies eq.\,(3.1), with $Z \eqq Z_1$;  $Z_1$
is unique by eq.\,(3.7) and is a central variable by eqs.\,(3.8), (3.9).

\noindent For a non compact $\M$, by using  Proposition 2.6, for  any $g \in
C^\infty(\M)$, $Z_g$ is unique, by eq.\,(3.7) with $h = 1$ in a region of
localization of both $Z_g$ and $Z'_g$. Moreover,  for any increasing sequence
of compact sets $K_n$ which cover $\M$, any sequence $g_n \in C^\infty(\M)$
with compact support and $g_n(x) = 1$, $\forall x \in K_n$, satisfies \be{ g_n
\cdot A = A \cdot g_n = A, \,\,\,\, n > \bar{n}(A),}\ee for all localized $A$,
as a consequence of Proposition 2.6. Then, putting $Z_n \eqq Z_{g_n}$,   for
$A$ localized in $K$ and $h$ such that $h \cdot A = A$, one has $$Z_n \cdot A =
Z_n \cdot h \cdot A = Z'_n \cdot h \cdot A = Z'_n \cdot A, $$ by eq.\,(3.7),
i.e., for $n$ large enough, $Z_n \cdot A$ is independent of $n$ and of the
construction of $Z_n$; again by eq.\,(3.7), $Z_n \cdot A$ is localized in $K$.
Moreover,  by eqs.\,(3.5), (3.8), (3.9),  the sequences $Z_n $ satisfy
eq.\,(3.2).
%\goodbreak

\noindent Hence,  one may consider the Poisson algebra $\tilde{\La}_R(\M)$
generated by $\LM$ and by an element $Z$ defined by $$Z \cdot A \eqq \lim_{n
\ra \infty} Z_n \cdot A = \lim_{n \ra \infty} A \cdot  Z_n \eqq A \cdot
Z,\,\,\,\, \mbox{for all localized}\,A,$$ $$Z \cdot \id \eqq Z \eqq \id \cdot
Z, \,\,\,\,\,\{ Z, \,A \} = 0, \,\,\,\,\,\forall A \in \LM.$$

\noindent Since, for localized $A$, $Z\cdot A$ is localized, $Z^n \cdot A \eqq
Z^{n - 1} \cdot (Z \cdot A)$ is recursively  well defined; associativity and
the Leibniz rule clearly hold and eq.\,(3.1) holds in $\tilde{\La}_R(\M)$.
$\hfill \square$

\vspace{1mm} In the above proof, conditions i)- iii) enter only after
eq.\,(3.9), hence
 \begin{Corollary} For the Poisson universal enveloping algebra $\La_{L(\M)}$
 of the Lie algebra
$L(\M) \eqq C^\infty(\M) + \L(\M)$ the commutators and the Lie products are
related by \be{ [\,A, \,B\,] \cdot g = \{\,A,\,B\,\} \cdot Z_g, \,\,\,\,\,A,
\,B \in \La_{L(\M)}, \,\,g \in C^\infty(\M), }\ee \be{ \{\,Z_g \cdot g, \,
A\,\} = 0 = [\,Z_g \cdot g,\,A\,], \,\,\,\,\,\,\,if \,\,\,\{\,A,\,g\,\} =
0.}\ee
\end{Corollary}
Furthermore, since only i) has been used  in the case of compact $\M$, one has
\begin{Corollary} For a compact  $\M$, Theorem 3.1 holds  for $\La(\M)$
= the Poisson UEA of the Lie algebra  $L(\M)$, satisfying condition i) of
Definition 2.4. \end{Corollary}

For the implications of the above results on Classical and Quantum Mechanics,
we note that a notion of reality in $\LM$ is automatically given  if the
elements of $C^\infty(\M) + \L(\M)$ are taken as real: \be{ f^* = f , \,\,\,v^*
= v.}\ee The {\em $^*$ operation} obviously extends to a (real linear)
involution on $\LM$, with $\{\,A,\,B\}^* = \{\,A^*,\,B^*\,\}$, $(A B)^* =
B^*\,A^*$ and therefore $Z^* = - Z$. Then, $\LM$ becomes a Poisson  *-algebra
(Definition 4.1 below) and as such will be understood in the following.

``{\em Factorial realizations}'' of $\LM$ are defined  by homomorphisms $\pi_z$
obtained by taking the quotient of $\LM$ with respect to the  ideal generated
by  $Z^* Z - z^2$, $z \geq 0$. (For Hilbert space representations see Section
4).

\noindent For $z = 0$, one obtains  the commutative Poisson algebra generated
by $\cifm$ and by the polynomials of the $C^\infty$ vector fields, with the
natural module structure of $\L(\M)$ on $\cifm$, i.e., substantially, the
symplectic structure of  classical mechanics (see Section 4) . Such an algebra
is also very close to the Lie algebroid advocated as the algebraic structure of
classical mechanics~\cite{La}.

\noindent For $z^2 \neq 0$, $\pi_z(Z) = \iota z$, $\iota^2 = -1$ and there is
an isomorphism $\ph$,  mapping the real Poisson involutive algebra into the
complex algebra generated by $C^\infty(\M)$ and by the generalized momenta
$T_v$ associated to the vector fields of $\L(\M)$, satisfying $$[\,T_v, \,T_w
\,] = i\, z\, T_{\{ v, \,w \}}, \,\,\,\,\,[T_v, \,f ] = i\,
 z\, \{ v , \, f\}, \,\,\,\,z \geq 0,$$ with
 $\ph(f) = f$, $\ph(v) = T_v$, $\ph(\iota) = i.$
This is the (unbounded) LR quantum algebra introduced in Ref.[1]. Such an
isomorphism between $\pi_z(\LM)$ and the above complex algebra offers an
explanation of the occurrence of a complex structure in the standard
formulation of quantum mechanics, whereas no complex structure is needed for
the formulation of classical mechanics.

In conclusion, the Poisson UEA $\LM$, with the reality notion (3.13), is common
to classical and quantum mechanics, which can be characterized as its only
factorial realizations, in the sense defined above.\goodbreak

\vspace{1mm}The central relation between commutators and Lie products, given by
Corollaries 3.2, 3.3, are independent of the LR relations and therefore apply
to other geometric structures which can be associated to $\difm$, without
including the LR relations. In fact, the Lie algebra $L(\M) = C^\infty(M) +
\L(M)$ can be interpreted as the current Lie algebra on $\M$ generated by the
charge densities $\rho(f)$, $f \in C^\infty(\M)$ and the currents $j(v)$, $v
\in \L(M)$, with \be{ \{\,\rho(f),\,\rho(g)\,\} = 0, \,\,\,\,\{\,j(v),
\,\rho(f)\,\} = \rho(\{ v, f\}), \,\,\,\,\{\,j(v), \,j(w)\,\} = j(\{ v,
\,w\,\}), }\ee for all $f, g \in C^\infty(\M), $ $v, \,w \in \L(\M)$. Hence,
$\La(\M)$ is identified with the Poisson UEA of the current Lie algebra with
the additional relation i) of Definition 2.4,  $i(\rho(1)) \cdot A = A$,
$\forall A \in \La_{L(\M)}$. Such a relation holds for the $N$-particle
representations of the current Lie algebra introduced by Goldin~\cite{G},
provided $\rho$ is identified with the Goldin charge density divided by $N$.

With the above identifications, the Poisson algebra $\La(\M)$ describes both
the classical and the quantum current algebra and for compact $\M$, by
Corollary 3.3, these are the only factorial realizations, in the above sense.
The LR relations are not satisfied in the current algebra, where in particular
the algebra generated by the $\rho(f)$ is a free abelian algebra.

\vspace{1mm} As it is clear from the proof of Theorem 3.1, the central variable
relation between commutators and Lie products follows in general from the
existence of an identity which can be written as a sum of Lie products. As a
physically relevant example, we consider the {\bf Lie algebra of the canonical
variables} in $\Rbf^n$, defined by the following Lie products \be{ \{\,q_i,
\,q_j \} = 0 = \{ p_i, \,p_j\,\}, \,\,\,\,\, \{\,q_i, \,p_j\,\} =
\delta_{i\,j}\,\id,\,\,\,\,\{ q_i, \, \id \} = 0 = \{ p_i, \, \id\},}\ee
henceforth denoted by $\L_c$. The construction of Theorem 3.1 simplifies and
the corresponding Poisson UEA $\La_c$, defined with the condition i) of
Definition 2.4, $i(\id) \cdot A = A$, contains a central variable $Z$, such
that $$ [ q_i, \,q_j ] = Z \{ q_i, \,q_j \} = 0,\,\,\,\,\,\,[ p_i, \,p_j ] = Z
\{ p_i, \,p_j \} = 0,$$ \be{[\,q_i, \,p_j \,] = Z \{\, q_i, \,p_j\,\} = Z
\delta_{i \,j}\,.}\ee Thus, the commutator between any pair of canonically
conjugated variables is a central variable in $\La_c$ and coincides with $Z$.

The Poisson algebra $\La_c$ describes both classical and quantum mechanics, the
two realizations corresponding, as above,  to the quotients of $\La_c$ with
respect to the  ideals generated by $Z^*\,Z - z^2$, $z \geq 0$,  with,
respectively,  $z = 0$ and  $z = \hbar$.

%%%%%%%%%%%%%%%%%%%%%%%%%%%%%%%%%%%%%%%%%%%%%%%%%%%%%%%%%%%%%%

%%%%%%%%%%%%%%%%%%%%%%%%%%%%%%%%%%%%%%%%%%%%%%%%%%%%%%%%%%%%%%%%%%

\def \cifm {C^\infty(\M)}
\def \difm  { \mbox{Diff}(\M)}
\def \gm {\G(\M)}
\def \gtm {\tilde{\G}(\M)}
\def \cmo {{C_0^\infty(\M)}}
\def \cif {{C^\infty}}
\def \co   {{C^\infty_0(\O)}}

\def \cst   {{C^*}}
\def \lm {{\cal L}(\M)}
\def \tv {T_v}
\def \lo {\L(\O)}
\def \ulv {U(\l v)}
\def \rv {R_v}
\def \rlv {R_{\l v}}
\def \rav {{R_{\a v}}}
\def \am {\A(\M)}

\def \ois {{\O \in \S}}
\def \glv  {{g(\l v)}}
\def \GM  {{\G(\M)}}
\def \Gtm  {{\tilde{\G}(\M)}}
\def \tcmo {{\tilde{C}^\infty_0(\M)}}
\def \CM  {\Pi(\M)}
\def \tilm {\tilde{\cal L}(\M)}
\def \ao {{\A(\O)}}

\def \LRM {\La_R(\M)}
\def \glzv {g_{\l z v}}
\def \Ulv {U(\l v)}
\def \ulv {\Ulv}
\def \umlv {U(- \l v)}
\def \ulz {U(\l Z)}
\def \rv {R_v}
\def \rw {R_w}
\def \rfv {R_{f \circ v}}
\def \glv {g_{\l v}}
\def \glw {g_{\l w}}
\def \lv {\l v}
\def \gotA {{\cal A}(\M)}

%%%%%%%%%%%%%%%%%%%%%%%%%%%%%%%%%%%%%%%%%%%%%%%%%%%%%%%%%%%%%%
\section{Representations of $\LM$. Classical and Quantum Mechanics}
The  Poisson algebra $\La_R(\M)$ involves unbounded variables and  the analysis
of its representations   is conveniently done in terms of associated normed
*-algebras and $C^*$-algebras. For this purpose, we introduce the following
notions.
\begin{Definition} A Poisson *-algebra $\La$ is a (noncommutative) Poisson
algebra, with a (real linear) involution *, satisfying $\{\,A,\,B\}^* =
\{\,A^*,\,B^*\,\}$, $(A \cdot  B)^* = B^*\cdot A^*$.

A {\bf representation} $\pi$ of a Poisson *-algebra $\La$ in a Hilbert space
$\H$ is a  homomorphism of $\La$ into a Poisson *-algebra of operators in $\H$
(with both the operator product and a Lie product $\{.\,, \,.\}$ satisfying the
Leibniz rule), having a common invariant dense domain $D$ on which  $$\pi(A
\cdot B) = \pi(A) \,\pi(B), \,\,\,\,\pi(\{\,A,\,B\,\}) =
\{\,\pi(A),\,\pi(B)\,\}, \,\,\,\,\pi(A^*) = \pi(A)^*.$$
\end{Definition}
\begin{Proposition} In a representation $\pi$ of (the Poisson *-algebra) $\LM$
in a  Hilbert space $\H$,
$\pi(f)$ and $\pi(v)$, $f \in \cifm$, $v \in \L(\M)$ are strongly continuous on
$D$ in the $C^\infty$ topology of $\cifm$ and $\L(\M)$.

Furthermore, eq.\,(2.6) holds for $\pi(\glv(A))$, $A = f,\, v$, with the
derivative taken in the strong topology.
\end{Proposition}
\Pf\,Since for $|\l| > ||f ||_\infty$, $f - \l $  is invertible in $\cifm$, $||
\pi(f)|| \leq ||f||_{\infty}$. Moreover,  any compact set $K \subset \M$ can be
covered by a finite number of open sets $\O$ homeomorphic to spheres, so that,
by a corresponding decomposition of unity, any vector field $v$ can be written
as a  linear combination $v = \sum_i^{d}\,f_i\,w_i$, $f_i \in \cifm$. Then,
$v_n \ra v$ in the $C^\infty$ topology implies that  the corresponding
$f_i^{(n)}$ converge  $f_i$ in $C^\infty(\O)$ and, by the LR relation $\pi(f_i
w_i) = \pi(f_i)\,\pi(w_i) + \pi(w_i(f_i)) Z$,  $\pi(v_n)$ converge strongly to
$\pi(v)$ on $D$. This also implies that $\pi(\glv(A))$ is strongly
differentiable in $\l$ and eq.\,(2.6) holds for $\pi(\glv(A))$.$\hfill \square$
\def \amw { g_{\mu w}}
\begin{Definition} A representation $\pi$ of $\LM$ in a complex Hilbert space
$\H$ is called {\bf regular} if $\pi(C_0^\infty(\M)) \neq 0$ and \newline i)
({\bf exponentiability}) $D$ is invariant under the $C^*$-algebra $\gotA_\pi$
generated by $\pi(\cifm)$ and by one parameter unitary groups $\ulv$, $\ulz $,
$\l \in \Rbf$, generated by $T_v \eqq \pi(v)$ and $- i\, \pi(Z)$, respectively,
\newline ii) ({\bf diffeomorphism invariance}) the elements $\alv \in \G(\M)$
define strongly continuous automorphisms of $\gotA_\pi$  by \be{\amw: \pi(f)
\ra \pi(\amw f),\,\,\,\, \ulv \ra U(\l \amw(v)), \,\,\,\,\,\ulz \ra \ulz.}\ee
\end{Definition}
Property ii) states the exponentiability of the derivation (2.6) in the
representation $\pi$; it is implied by i)  if $z \neq 0$ (see below).
\begin{Proposition} In a regular representation $\pi$ of $\LM)$, the
  one-parameter unitary groups
$\ulv$, $\ulz $ satisfy \be{ [\,\ulv, \,\ulz\,] = 0, \,\,\,\,\,\,[\,\pi(f),
\,U(\l Z)\,] =0.}\ee
\end{Proposition} \Pf\,\,In fact, one has on $D$ $$i (d/ d \l)[ \ulv
\,\pi(Z)\,\umlv\,] = \ulv \, \,[\, T_v,\,\pi(Z)\,]\,\umlv = 0,$$ i.e. $[\,\ulv,
\,\pi(Z)  \,] = 0$; therefore $(d/d \l)\,\ulz\, U(\mu v) \,\ulz^{-1} = 0$.
Similarly for $\pi(f)$ and eqs.\,(4.2) follow. $\hfill \square$

Since $\LM$ has both an associative product and a Lie product related to
$\difm$ by eq.\,(2.6), a natural role  is played by  elements which are central
with respect to both.
 \begin{Definition} A
regular representation $\pi$ of $\LM$ is called {\bf factorial} if the elements
of  the center ${\cal Z}_\pi$ of the Von Neumann algebra $\gotA_\pi''$ which
are invariant under $\difm$ are multiples of the identity.
\end{Definition}

Classical and Quantum Mechanics are examples of regular (factorial)
representations of $\LM$. \vspace{1mm} \newline A) {\bf Quantum Mechanics}. As
discussed in Ref.\,[1], QM on a manifold  $\M$, with $\hbar = 1$, is described
by the {\em LR regular} representations $\pi$ of the crossed product $\Pi(\M)
\eqq \cifm \times \Gtm$, i.e. such that $\pi(C_0^\infty(\M)) \neq 0$ and
\newline i) $\pi(U(\l v))$ are strongly continuous in $\l$,
\newline ii) their generators $T_v$ have a common dense domain $D$ invariant
under $\Pi(\M)$, satisfy the Lie algebra relations and the LR relations,
$\forall f \in C^\infty(\M), \,\,v, w\, \in \L(\M)$,  $$ [\,T_v, \,T_w\,] = i
\,T_{\{v, \,w\}},\,\,\,\,\,\,\, T_{f \circ v} = \ume (f\,T_v + T_v\,f).$$ By
Propositions 4.2, 4.6, $\pi(\ulv)$ are also strongly continuous in the
$C^\infty$ topology of the vector fields. The classification of irreducible
such representations~\cite{MS1}, implies, by their locally Schroedinger
property, that $D$ can be chosen invariant under the generators $T_v$ and
therefore one actually gets a regular representation of $\LM$ with $Z = i$. The
automorphisms $\glv$ are implemented by $\ulv$ and property ii) of  Definition
4.3 follows; the representation of $\LM$ is factorial if $\pi$ is irreducible.

More generally, the above  operators $f$ and $T_{\hbar v} $ define  regular
representations of $\LM$ with $Z = i \,\hbar$. Such representations are
therefore classified by $\hbar$ and the unitary representations of the
fundamental group $\pi_1(\M)$ ~\cite{MS1}; they shall be called the {\em
quantum representations} of $\LM$.

\vspace{1mm}\noindent B) {\bf Classical Mechanics}. Given the cotangent bundle
$T^*\M$, described by local coordinates $(x, p)$, with $x \in \M$ and $p_i$ the
coordinates in the basis dual to $ \partial/\partial x_i$, we consider the
Hilbert space $L^2(T^* \M, d x \,d p)$ and the representation $\pi_C$ of $\LM$,
with $Z = 0$, by multiplication operators on $D = C_0^\infty(T^* \M)$, $\forall
f \in \cifm$, $\forall v = \sum_i g_i(x)
\partial/\partial x_i$, supp\,$v \subseteq \O$, ($\forall\O$
homeomorphic to  an open  sphere),  $$ \pi_C(f) = f(x), \,\,\,
\,\,\,\,\,\,\pi_C(v) = \sum_i g_i(x)\, p^i \eqq T_v, \,\,\,\,\,\ulv = e^{- i \l
\sum_i g_i(x)\,p^i}.$$ The Lie product on $\LM$ is given by the standard
Poisson brackets $\{\,.\,,\,.\,\}_C$ on $T^* \M$. The automorphisms $\glv$,
$\glv(f)(x) = f(\glv^{-1}(x))$, $\glw(T_v) = T_{\glw(v)}$, satisfy eq.\,(2.6),
extend to the $\ulv$ and are unitarily represented. Hence, $\pi_C$ is a regular
representation of $\LM$, (actually it is factorial), with $Z = 0$. It shall be
referred to as the {\em classical canonical representation} of $\LM$.

In the following, we show that the regular factorial representations of $\LM$
are quasi equivalent  to either one of the quantum irreducible representations
or (in separable Hilbert spaces) to  the classical canonical  representation.

\begin{Proposition} In a regular factorial representation $\pi$ of $\LM$
one has \newline i) $\ulz = e^{- i \l z} \id $, $z \in \Rbf$; apart from an
antiunitary transformation leaving $\LM$ pointwise invariant, one can take $ z
\geq 0$, \newline ii) the one parameter groups $\ulv$ are strongly continuous
in $v$ in the $C^\infty$ topology of the vector fields and
 \be{U(\l v)\,U(\mu\,w) = U(\mu g_{\l z v}(w)) \,U(\l v), \,\,\,\,\,\,\,\,
 U(\l v) f  = \glzv(f)\,U(\l v), }\ee  $\glzv( \cdot )$ denoting  the action of the one
parameter group of $\G(\M)$ generated by $\l z v$.
\end{Proposition}
\Pf\,\, By eq.\,(4.2), $U(\l Z)$ belongs to ${\cal Z}_\pi$ and, by eq.\,(4.1),
is invariant under $\difm$; then, $U(\l Z) = e^{-i \l z} \id$, by strong
continuity in $\l$. For the antiunitary transformation see Section 3.

\noindent Moreover, by using Proposition 4.2, $\forall A \in C^\infty(\M) +
\L(\M)$ one has $$(d/d \l)\,[U(-\l v)\,\pi(\glzv(A))\,\ulv\,] = $$ $$= \umlv\,
\big{(}\,i\, [\,T_v, \,\pi(\glzv(A))\,] + \{\, z T_v, \,\pi(\glzv(A))\}
\,\big{)}\, \ulv = 0.$$ Then, $$ \umlv\,\pi(\glzv(A))\,\ulv = A$$ and
eqs.\,(4.3) follow, by exponentiability. Finally, $\forall \Psi \in D$,
eq.\,(4.3) implies $$(d/d \l) \Psi_n(\l) \eqq (d/d\l)
 [U(\l v_n)\,U(-\l v)]\,\Psi = U(\l v_n)\,
U(- \l v) (T_{g_{- \l z v}(v_n)} - T_v)\Psi $$ and $T_{g_{- \l z v}(v_n)} \Psi
= T_{g_{- \l z v}\,g_{\l z v_n}(v_n)} \Psi$ converges strongly to $T_v \Psi$,
uniformly in $\l$ (bounded), by Proposition 4.2; hence, $\Psi_n(\l) = \Psi +
\int_0^\l d \l'\,(d/d \l') \Psi_n(\l') \ra \Psi.$

$\hfill \square$
\begin{Theorem} The regular factorial representations  $\pi$ of $\LM$ are
classified by the value  $i\, z$ of the central variable $Z$ and  \vspace{1mm}
\newline 1) for $z \neq 0$, each of them is   unitarily equivalent, apart from a
multiplicity,  to one of the irreducible {\bf quantum representations} defined
above by a LR regular irreducible representation of the crossed product $ \cifm
\times \Gtm$, with $z = \hbar$, and therefore in one to one correspondence with
the unitary representations of $\pi_1(\M)$ \vspace{1mm} \newline 2) for z = 0,
for separable representation space $\H$, they are quasi equivalent to the {\bf
classical canonical representation}, actually, unitarily equivalent, up to a
multiplicity, if $\difm$ is unitarily implemented.
\end{Theorem}
\Pf\,\,\,For $z \neq 0$, by eqs.\,(4.3), $\pi(f)$, $f \in \cifm$ and the $U(\l
z^{-1} v)$, $v \in \L(\M)$,  provide a representation of the crossed product
$\Pi(\M)$, which is LR regular by Definition 4.3, i).

\noindent Since the automorphisms $\glv$ are unitarily implemented by the $U(\l
z^{-1} v)$, the center of $\Pi(\M)''$ is automatically pointwise invariant
under $\G(\M)$ and, therefore, factoriality in the sense of Definition 4.5
implies factoriality of the representation of the crossed product. Hence, the
representation is a sum of copies of one irreducible representation of
$\Pi(\M)$; the latter are classified by the unitary irreducible representation
of $\pi_1(\M)$, Ref.\,[1] . As remarked before, each LR regular irreducible
representation of $\Pi(\M)$ defines a regular factorial representation of
$\LM$.

\def \gota {{\cal A}}
\def \gotap {\gotA_\pi}
\def \pxi {p^i_\xi}
\noindent For $z = 0$, by eqs.\,(4.3), the $C^*$-algebra $\gota(\M)_\pi$
generated  by $\pi(\cifm)$ and the $\ulv$ is abelian; by  separability, the
representation space can therefore be identified with   a sum  of $L^2$ spaces
with finite measures $\sigma_n$ over the spectrum $\Sigma$ of $\gotap$, whose
points $\xi$, apart from sets of zero measure for all $\s_n$, are identified as
multiplicative linear functionals which by regularity can be characterized as
follows: $\forall f \in \cifm$, $\xi(f) = f(x_\xi)$, and  $\forall v = \sum_i
g_i(x)
\partial/\partial x_i$, supp\,$v \subseteq \O$, ($\forall\O$
homeomorphic to  an open  sphere) $$\xi(v) = \sum_i g_i(x_\xi) \,p^i_
\xi,\,\,\,\,\pxi \eqq \xi(\partial/\partial x_i), \,\,\,\xi(\ulv) = \exp{( i
\sum_i g_i(x_\xi)\,\pxi)},$$ i.e. the point $\xi$ is identified by local
coordinates $(x, p)$ in  $T^* \M$. Then, apart from a set of zero $\s_n$
measure, $\Sigma$ can be identified with a subset of $T^* \M$,  by a continuous
map, so that $\s_n$ defines a finite Borel measure $\mu_n$ on $T^* \M$ and the
representation space $\H$ may be taken as $\sum_n L^2(T^* \M, d \mu_n)$.

\def \tstm {T^* \M}
\def \glvi {g_{\l_i\,v_i}}
\noindent The set of points $ I_0 \eqq \{(x, 0), \,x \in \M \}$ is invariant
under $\difm$ and it has zero $\mu_n$-measure, since, otherwise, it would
define a subrepresentation of $\LM$, which violates the validity of eq.\,(2.6),
as stated by Proposition 4.2. The action of $\difm$ on $\tstm$ is transitive on
the points $(x, p)$, $p \neq 0$; moreover, for each point $(x, \,p),\, p \neq
0$, there are $2 d$ vector fields $v_i$, such that $\l_i$, $i= 1, ...2 d$,
$|\l_i| < 2 \eps$, define local coordinates for $\glvi(x, p)$. If $I \subseteq
\tstm$ is a Borel set of zero Lebesgue measure, contained in  a neighborhood of
$(\bar{x}, \bar{p})$,  one has (by Fubini theorem) $$ \int_{ |\l_i| < \eps} \Pi
d \l_i \int d \mu\, \chi_{I,\, \glvi} =\int d \mu \int_{ |\l_i| < \eps} \Pi d
\l_i \,\chi_{I,\, \glvi} = 0,\,\,\,\,\,d \mu \eqq \sum_n 2^{-n}\,d \mu_n,$$
where $\chi_I$ is the characteristic function of $I$ and $\chi_{I, \glvi}(x, p)
\eqq \chi_I(\glvi^{-1}(x, p))$. Then, $\int d \mu\, \chi_I(\glvi^{-1}(x, p)) =
0$, a.e. in $\l_i$ and, since the projection $E_I$ and $\glv(E_I) =
E_{\glv^{-1}(I)}$ are isomorphic, $\int d \mu\, \chi_I = 0$. Therefore, one has
a  representation of $\LM$, as an abelian algebra,  quasi equivalent to the
classical canonical representation. Conversely, if $\mu(I) = 0$ and $I$ has non
zero Lebesgue measure, $\int_{|\l_i| < \eps} \Pi d \l_i\,\chi_{I, \,\glvi}(x,
p) > \delta > 0,$ for $(x, p)$ in an open set $\O_I$ and therefore the above
equation implies $\mu(\O_I) = 0$; by the transitivity of $\difm$, $d \mu = 0$.
Therefore, $d \mu$ is equivalent to the Lebesgue measure.

\noindent The Lie product on $\pi(\LM)$ is completely determined by its
restriction to $C^\infty(\M) + \L(\M)$, which, by Proposition 4.2  satisfies
eq.\,(2.6) and locally one has, $\forall A(x, p) \in C^\infty(\M) + \L(\M)$, $$
\{\sum_i g_i(x) \,p^i,\,A(x, p)\,\} = (d/ d \l) A(\glv^{-1}(x, p))|_{\l = 0} =
$$ $$ = \sum_i \Big{(} - \frac{\partial A(x, p)}{\partial x_i}\, g_i(x) +
\frac{\partial A(x, p)}{\partial p^i}\,\frac{\partial g_j(x)}{\partial x_i}\,
p^j \Big{)} = \{\sum_i g_i(x) \,p^i,\,A(x, p)\,\}_C.$$ Thus, the representation
of $\LM$, as Poisson algebra, is given by the classical canonical
representation, up to multiplicities. If $\difm$ is unitarily implemented, the
multiplicity of the representation is $\mu$ a.e. constant and therefore it is a
multiple of the classical canonical representation. $\hfill \square$

\end{document}